\documentclass[twocolumn]{aastex631}
\usepackage[]{amsmath}

\begin{document}

\title{Planetesimal Growth in Evolving Protoplanetary Disks: Constraints from the Pebble Supply }

\correspondingauthor{Tong Fang}
\email{tfang@shao.ac.cn}

\author{Tong Fang}
\affiliation{Shanghai Astronomical Observatory, Chinese Academy of Science, People's Republic of China}
\author[0000-0003-3491-6394]{Hui Zhang}
\affiliation{Shanghai Astronomical Observatory, Chinese Academy of Science, People's Republic of China}
\author[0000-0002-9442-137X]{Shangfei Liu}
\affiliation{School of Physics and Astronomy, Sun Yat-sen University, Zhuhai, People's Republic of China}
\author[0000-0001-5830-3619]{Beibei Liu}
\affiliation{Zhejiang Institute of Modern Physics, Department of Physics and Zhejiang University–Purple Mountain Observatory Joint Research Center for Astronomy, Zhejiang University, 38 Zheda Road, Hangzhou 310027, People's Republic of China}
\author[0000-0001-6858-1006]{Hongping Deng}
\affiliation{Shanghai Astronomical Observatory, Chinese Academy of Science, People's Republic of China}



\begin{abstract}
In the core accretion model, planetesimals grow by mutual collisions and engulfing millimeter-to-centimeter particles, i.e., pebbles. Pebble accretion can significantly increase the accretion efficiency and help explain the presence of planets on wide orbits. However, the pebble supply is typically parameterized as a coherent pebble mass flux, sometimes being constant in space and time. Here we solve the dust advection and diffusion within viciously evolving protoplanetary disks to determine the pebble supply self-consistently. The pebbles are then accreted by planetesimals interacting with the gas disk via gas drag and gravitational torque. The pebble supply is variable with space and decays with time quickly, with a pebble flux below 10 $M_\oplus$ Myr$^{-1}$ after 1 Myr in our models. As a result, only when massive planetesimals ($>$ 0.01 $M_\oplus$) are luckily produced by the streaming instability or the disk has low viscosity ($\alpha \sim 0.0001$) can the herd of planetesimals grow over a Mars mass within  2 Myr. By then, planetesimals only capture pebbles about 50 times their mass and as little as 10 times beyond 20 au due to limited pebble supply.  Further studies considering multiple dust species in various disk conditions are warranted to fully assess the realistic pebble supply and its influence on planetesimal growth.
\end{abstract}



\section{Introduction} \label{sec:intro}

In the core accretion model of planet formation, dust in protoplanetary disks coagulates to form progressively massive bodies, i.e., pebbles, planetesimals, protoplanets/planetary embryos (here we refer to objects more massive than Mars) , and planetary cores, which may further capture gas to form giant planets \citep{Pollack1996,Liu2020,drazkowska2022}.  Early dust growth to millimeter size is witnessed in young disks \citep{facchini2019,galametz2019,Segura2020}. The protoplanetary disk retains these dust particles, termed pebbles, for at least several million years \citep{williams2011,andrews2016}. However, pebbles need to be converted to planetesimals with significant inertia; otherwise, they are expected to drift toward the central star and get accreted quickly \citep{Weidenschilling1977}.

The formation and growth of planetesimals are arguably the least known part of the core accretion theory \citep{Chiang2010}. The streaming instability can lead to dust clumping and, subsequently, the collapse of dust clumps to form planetesimals providing a high initial dust-to-gas ratio near unity \citep{Youdin2005}. However, the required high dust content \citep[but see][, and references therein]{Li2021} is rarely found in protoplanetary disks except at special regions, such as snow lines \citep{Drkazkowska2017,Schoonenberg2018} and pressure bumps \citep{Pinilla2017,Miller2021}. Alternatively, \cite{Tominaga2021} proposed an instability (albeit slow) due to the interplay between dust-coagulation efficiency and dust radial drift to enhance the dust-to-gas ratio and set the stage for planetesimal formation via the streaming instability or secular gravitational instability  \citep{Pierens2021}.

Planetesimals with diameters of hundreds of kilometers can grow in two ways: mutual collisions and engulfing pebbles from the dusty disk. Accretion through mutual collisions can result in runaway growth when more massive planetesimals grow at a faster rate \citep{greenberg1978,wetherill1989}. Subsequently, planetary embryos grow oligarchically while the rest planetesimals remain small \citep{Kokubo1998}.  On the other hand, planetesimals can accrete pebbles with the assistance of aerodynamic drags. Pebble accretion was envisaged by \cite{Ormel2010} and \cite{Lambrechts2012} and is believed to overtake collisional growth in the late stage of planetesimal mass growth under the nominal disk conditions \citep{liu2019growth}. Pebble accretion is crucial for building up the cores of giant planets within the lifetime of the gas disk \citep{Lambrechts2014}. Nevertheless, pebble accretion may still be inefficient for massive planets beyond 100 au \citep{Bae2022} and around low-mass stars \citep{Morales2019,Liu2020pebble} for which disk instability may be a viable formation channel \citep{Deng2021}.

The pebble accretion efficiency depends on the pebble supply and size, and the orbit of the planetesimal \citep{Liu2018}. Typically, the pebble supply is provided by a global pebble flux \citep[see,e.g.][]{Lambrechts2019,Ogihara2020,Liu2022} and sometimes a moving pebble production line \citep{Lambrechts2014,Izidoro2021}. The pebble flux is found to be decisive in the resultant planet population, where massive planets are common given a high pebble flux \citep{Lambrechts2019}.  However, the assumption of such a coherent pebble flux being radius and/or time-independent, may not hold for general protoplanetary disks \citep{lin2018balanced}. In disks, the dust evolves by an advection-diffusion equation \citep{Clarke1988,Birnstiel2010,Zhou2022}, and nontrivial dust motion may happen, including layered upstream diffusion \citep{Zhou2022} and piling up as dust rings \citep{Tominaga2022}. 

Here we assess how the variation of the pebble supply in space and time affects planetesimal accretion. We do so by coupling $N$-body simulations of planetesimal evolution to pebble accretion subject to self-consistent pebble drift and diffusion \citep{Zhou2022}. We introduce our model in Section \ref{sec:method}, including the pebble supply, planetesimal evolution, and simulation setup. Next, we present our simulation results in Section \ref{sec:results} and discuss their implications for planet formation in Section \ref{sec:discussion}. Finally, we conclude in Section \ref{sec:conclusions}.

\section{Method and Numerical Models}

\label{sec:method}
\begin{figure}[ht!]
\plotone{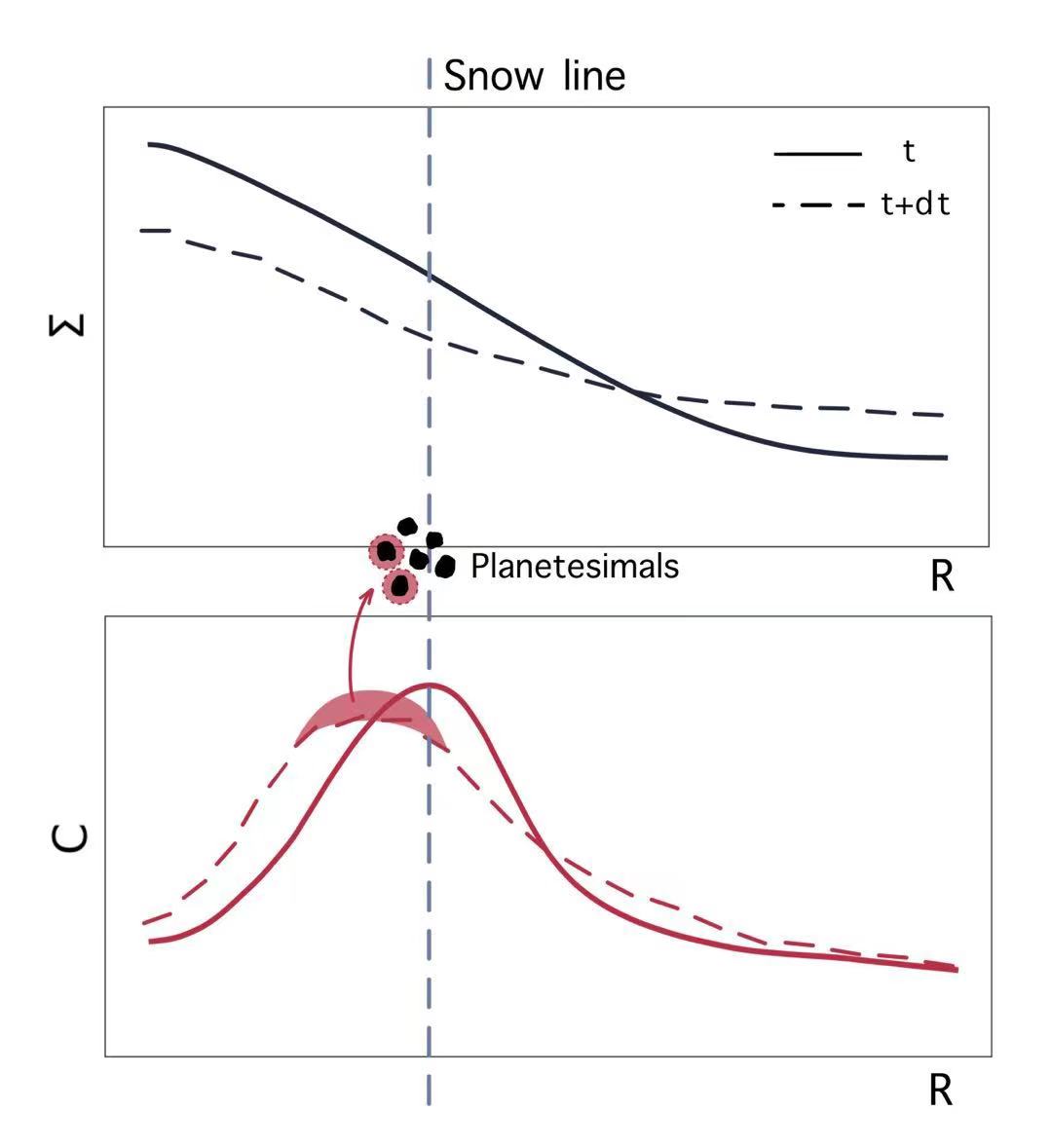}
\caption{Schematic plot of the dusty disk and planetesimal coevolution. Here $\Sigma$ and $C$ are the gas surface density and dust concentration, respectively. As the dusty disk evolves, the planetesimals accrete dust (red shaded regions) and migrate in the disk. Here $dt$ is the time interval per dusty disk status update, and it is much longer than the $N$-body time step (see the main text).\label{fig:illustration}}
\end{figure}

We intend to model the planetesimal accretion in an evolving disk self-consistently. Planetesimals may preferentially form near the snow lines and then migrate and accrete \citep{Drkazkowska2017,liu2019growth}. The orbital evolution of planetesimals is simulated with the public REBOUND code \citep{Rein2012}. In the meantime, the dusty disk evolves and feeds the planetesimals with pebbles \citep{Stammler2022, Zhou2022}. This is illustrated in Figure \ref{fig:illustration}, showing the temporal evolution of the gas surface density ($\Sigma$) and dust concentration (dust-to-gas mass ratio). Notably, some pebbles are accreted from the dusty disk by the planetesimals (red shaded regions in Figure \ref{fig:illustration}), which further decreases the local dust concentration and total pebble mass. The details of the dusty disk and planetesimal coevolution are presented in the following subsections.

During the preparation of the paper, \cite{Lau2022} adopted a similar approach to coevolve the dust and planetesimal disk and find efficient planetesimal growth in stationary pressure bumps. In contrast, we focus on smooth disks without imposed disk substructures. We find inefficient planetesimal growth, which to some extent necessitates the presence of disk substructures for efficient pebble accretion.

\subsection{Pebble Supply} \label{sec:flux}
We adopt the standard viscous disk model, where a fraction of gas carries away most angular momentum and thus enables disk accretion \citep{Lynden1974}. The gas brings with itself dust particles leading to either inward or outward transport of pebbles. In addition, dust diffusion due to turbulent mixing can also alter the pebble transport in disks \citep{Clarke1988,Zhou2022}. The equations governing the evolution of the dust concentration $C$, which is defined as the dust surface density ($\Sigma_d$) to gas surface density ($\Sigma$) ratio, are \citep{Zhou2022}
\begin{align}
\Sigma \frac{\partial C}{\partial t}  = & \frac{1}{R}\frac{\partial}{\partial R}\left( R D \Sigma\frac{\partial C}{\partial R} \right) \nonumber \\
&- \frac{1}{R}\frac{\partial}{\partial R}\left( R \Sigma v_{\rm d} C \right)
+ C \frac{1}{R}\frac{\partial}{\partial R}\left(R\Sigma v_g\right)\label{eqn:dust-1layer},\\
\frac{\partial \Sigma}{\partial t}  = & \frac{3}{R} \frac{\partial}{\partial R}\left( R^{1/2} \frac{\partial}{\partial R} ( \Sigma \nu R^{1/2}) \right)\label{eqn:dusty}.
\end{align}
Here $ v_d $ and $ v_g $ are the dust and gas radial velocity, respectively. The kinetic viscosity, $ \nu = \alpha h^2 v_k R$, follows the $\alpha$ parameterization of \cite{Shakura1973}, where $ v_k = \sqrt{GM/R} $ is the Keplerian velocity, and $h$ is the disk's aspect ratio. The radial dust diffusivity is assumed to be $ D=D_0 h^2  v_kR $, similar to the viscosity \citep[see, e.g.,][]{Zhou2022}.
The gas radial velocity reads \begin{align}
v_g =&-\frac{3}{\Sigma R^{1/2}}\frac{\partial}{\partial R}(\nu \Sigma R^{1/2}),\\
=& - \frac{3\nu}{R}\left(\frac{\partial \ln\Sigma}{\partial \ln R}+\frac{\partial \ln\nu}{\partial \ln R}+\frac{1}{2}\right),
\end{align}
and the dust radial velocity is related to it by\begin{align}
v_d =\frac{v_g/T_s-2\eta v_k} {T_s+1/T_s},
\end{align}
where $ T_s $ is the Stokes number, and $\eta =-0.5 h^2 d\ln P /d \ln R$  measures the degree of sub-Keplerian rotation due to the gas pressure \citep{Takeuchi2002}. We note that equivalent equations for $\Sigma$ and $\Sigma_d$ are solved in studies of dust evolution \citep[see, e.g.,][]{Stadler2022, Stammler2022}. The $ T_s $ is closely related to the dust size, and \cite{Stammler2022} found a typical $T_s$ of a few thousandths (their Figure 11). Here, for simplicity, we consider a constant $T_s=0.001$ and leave a more comprehensive study with evolving $T_s$ to the future.

\subsection{Planetesimal Evolution}
\label{sec:pltacr}

The orbital motion of planetesimals is simulated with the public $N$-body code REBOUND \citep{Rein2012} with the Mercurius integrator, a hybrid of the WHFast and IAS15 integrator to capture collisions \citep{Rein2019}. Planetesimals are subject to gas drags \citep{Adachi1976}, and planet embryos also feel drags by spiral density waves excited by themselves and undergo type I migration. We follow the type I migration prescription of \cite{cresswell2008}, which is widely used in $N$-body simulations \citep[see, e.g.,][]{Lambrechts2019,liu2019growth,Ogihara2020,Izidoro2021,Jiang2022} as implemented in the REBOUNDx package \citep[][Kajtazi, K. et al. in preparation]{Tamayo2020}. 

We modified the REBOUNDx package to include also gas drags \citep{Adachi1976} for planetesimals and allow evolving disk surface density in the migration timescale calculation \citep{cresswell2008}. The acceleration due to gas drags reads
\begin{equation}
\pmb{\alpha}_\mathbf{drag} = -\left(\frac{3C_D\rho_\text{gas}}{8R_p\rho_p}\right)v_\text{rel}\pmb{v}_\mathbf{rel},
\end{equation}where $C_D=0.5$ is the nondimensional drag coefficient, and $v_\text{rel}$ is the relative velocity between the planetesimal and the gas \citep[see also][]{liu2019growth}. The density of the planetesimals, $\rho_p$, is set to 1.5 g cm$^{-3}$ and their radii, $R_p$, can be calculated accordingly. The accelerations due to type I migration are \citep{Papaloizou2000}
\begin{equation}
\pmb{\alpha}_\mathbf{m} = -\frac{\pmb{v}}{t_m}, \pmb{\alpha}_\mathbf{e} = -2\frac{(\pmb{v}\cdot \pmb{r}) \pmb{r}}{r^2 t_e}, \pmb{\alpha}_\mathbf{i} = -\frac{\pmb{v}_\mathbf{z}}{t_i},
\end{equation}
where the orbital, eccentricity, and inclination damping time scales, $t_m, t_e$, and $t_i$, are decided by the disk property \citep{cresswell2008}.

We note that the disk evolves on a viscous time scale much larger than the orbital periods of planetesimals. For example, an initial time step of several days is needed to simulate terrestrial planet formation in the inner solar system, let alone close encounters and collisions, which decrease the local time steps significantly \citep[see, e.g.,][]{Fang2020}. As a result, we can safely assume that the disk evolves little within a short period of $dt$ and calculate the above accelerations with fixed gas density.  We also apply this trick (a fixed $\Sigma_d$ over a short time interval) in the pebble accretion discussed below. We take a $dt$ of 10 yr in our simulation. We note that identical results are obtained should we update the disk status (according to section \ref{sec:flux}) three times more frequently, i.e., $dt=3$ yr.

We follow the basic pebble accretion recipe of \cite{Morbidelli2015} \citep[; see also][]{Ogihara2020}, and a more comprehensive treatment considering planetesimal eccentricities and inclinations can be found in \citet{Liu2018}. The effective radius for pebble accretion onto a planetesimal is
\begin{align}
r_\text{eff} =A R_\text{GP},
\end{align}
where $ R_\text{GP} = \text{min}(R_\text{B}, R_\text{H}) $ is the effective distance within the planetesimal's gravitational pull, with $ R_\text{B} $ being the Bondi radius and $ R_\text{H} $  being the Hill radius. The prefactor is found to be $A=(T_s/0.1)^{1/3}=0.22$ in the Hill accretion regime \citep{Lambrechts2012}.  In the Bondi accretion regime, $A=\sqrt{\eta^3 T_s /q}$ for a well-coupled dust species, where $q$ is the planetesimal-to-star mass ratio \citep[see][equation 27]{Lambrechts2012}. This formula leads to $A$  in a range of 0.03-0.13 for the models considered here (see section \ref{sec:setup}). However, polydisperse pebble accretion can result in 1-2 orders higher accretion rates than the single-species pebble accretion in the Bondi regime \citep{lyra2023}.  Here we simply take $A=0.22$ for the Bondi regime accretion as well. 

The pebbles are likely vertically extended with a scale height of $ H_\text{pb} = \sqrt{\alpha/(\alpha +T_s)} H $ \citep{Youdin2007}. When $H_\text{pb}$ is smaller than $ r_\text{eff} $, the accretion is 2D. The loading of pebbles happens across a line of width $2r_\text{eff}$. However, if $H_\text{pb} >  r_\text{eff}$, the accretion occurs in 3D with a cross section of $\pi r^2_\text{eff}$. The pebble accretion rates in the 2D and 3D regimes are
\begin{align}
\dot M_\text{2D} = 2 r_\text{eff} v_\text{acc} \Sigma_\text{d},\\
\dot M_\text{3D} = \pi r_\text{eff}^2 v_\text{acc} \rho_\text{d},
\end{align}
respectively. Here $ v_\text{acc} = \eta v_k + r_\text{eff}\Omega_k $ is the relative velocity between the pebbles and the planetesimal, with $\Omega_k$ denoting the Keplerian frequency \citep[see, e.g.,][]{Lambrechts2012,Ogihara2020}. The pebble flux $\dot{M}_\text{peb}=2\pi R \Sigma C v_d$ is often assumed to be a constant throughout the disk and used to derive the $\Sigma_d$ needed for pebble accretion. However, we directly obtain $\Sigma_d$ from solving the dusty disk evolution, so that $\dot{M}_\text{peb}$  is not used in our calculation but only given as a reference below. 

As noted above, the dust surface density is updated every 10 yr (i.e., $dt$ in Figure \ref{fig:illustration} with exaggerated changes) when we also subtract the amount of pebbles accreted by planetesimals during this time interval. The subtraction of planetesimal-accreted solids is done by simply decreasing the dust density in the cell containing the corresponding planetesimal.  That mass is then added to the planetesimal (smaller than $10^{-5}$ of its mass) perturbing the $N$-body system slightly. After each mass exchange between the dusty disk and planetesimals, the two systems are integrated separately with time steps borne by each system.

\subsection{Simulation Setup}
\label{sec:setup}
\begin{deluxetable*}{c|c|c|c|c|c|c|c|c|c|c}[ht!]
  \tablenum{1}
  \tablecaption{simulation parameters and results}
  \tablewidth{\textwidth}
  \tablehead{
    \colhead{Run } & \colhead{$M_\text{plt,0}$} & \colhead{$\alpha$}& \colhead{$D_0$} &\colhead{$R_\text{ice}$}  & \colhead{$C_1$} & \colhead{$\sigma$}&\colhead{$N_\text{c,all}$}& \colhead{$N_\text{c, max}$}& \colhead{$M_\text{max}$} & \colhead{$M_\text{peb}/M_\text{plt,0}$} }
  \startdata
  Fiducial &0.018     & 0.001     & $\alpha$& 2.7        &0.2  & 0.5       &22   &1 &0.023   &52.94   \\
  DustCon-1&0.018      & 0.001     & $\alpha$ & 2.7            &0.1      & 0.5     &25 &1   &0.014    &30.35  \\ 
  DustCon-2 &0.018      &  0.001    & $\alpha$    &  2.7       &0.3     &  0.5     &24  &7  &0.094    &87.49  \\
  FocusCon &0.018     & 0.001     & $\alpha$ & 2.7                  &0.2    & 0.1   &28  &0   &0.008  &18.88  \\ 
  DustDiff-1  &0.018       & 0.001     & 0 & 2.7           &0.2     &0.5        &36 &8   &0.021  &30.12   \\ 
  DustDiff-2  &0.018    &  0.001    & 2$\alpha$    &  2.7    &0.2   & 0.5        &22 &3  &0.026   &60.71  \\ 
  SnowLine-1 &0.115     & 0.001     & $\alpha$ & 20        &0.2      &0.5         &11 &2  &0.021  &7.25  \\ 
  SnowLine-2  &0.145   & 0.001     & $\alpha$ & 30        &0.2      &0.5         &10 &1  &0.023  &6.45\\   
 Test1 &0.018  &  0.0001   & $\alpha$    &  2.7       &0.2     & 0.5      &40 &2   &0.340      &627.01   \\ 
  Test2    &0.018   & 0.001     & $\alpha$ & 2.7          &0.2      &0.5        &29 &6 &0.098  &37.10 \\
  Test3  &0.018    & 0.001     & $\alpha$ & 2.7        &0.2     &0.5         &30 &6  &0.049  &52.87  \\
  \enddata
  \caption{The simulations include one fiducial model and others investigating the effects of dust concentration near the snow line, gas viscosity, dust diffusivity, and the location of the snow line. The Test1 simulation is identical to Fiducial but for lower viscosity. The other two models, Test2 and Test3, are similar to Fiducial but added with one massive planetesimal of 0.01 $M_\oplus$ and a Jupiter-mass planet at 5 au, respectively. Here $M_\text{plt,0}$ is the initial total mass of all planetesimals in $M_\oplus$; $M_\text{max}, M_\text{peb}$  are the mass of the largest planetary embryo and all accreted pebbles by the end of the simulations (2 Myr). The total number of collisions and the number of collisions involving the largest embryo are denoted as $N_\text{c, all}, N_\text{c,max}$, respectively.  \label{table:simulations}} 
\end{deluxetable*}

The formation of planetesimals is typically attributed to the streaming instability, which preferentially occurs at regions with enhanced dust concentration, $C$ \citep[cf.][and references therein]{Li2021}. This is still under investigation in evolving protoplanetary disks \citep[see, e.g.,][]{Lenz2019} and likely occurs in disks older than 1 Myr \citep{Estrada2023}. We adopt initial gas profiles similar to \cite{liu2019growth}, where
\begin{align}
\Sigma=&\Sigma_0 \left (\frac{R}{1  \text{au} }\right )^{-1},\\
h=& h_0 \left (\frac{R}{1  \text{au} }\right )^{1/4}.
\end{align}
Here we chose $ \Sigma_0 = 500$ g cm$^-2$ and $ h_0 = 0.033$, the same as \citep{liu2019growth}.   We discuss the effect of starting with a massive young disk in section \ref{sec:discussion}. We consider a simple yet effective model where planetesimals first form near snow lines at $R_\text{ice}$ \citep{Drkazkowska2017}. The water-ice snow line lies at about 2.7 au (default value), while the CO-ice line in protoplanetary disks ranges from tens of au to beyond 100 au \citep{Mathews2013,Qi2015chemical,Van2017}. We studied models with ice lines at 2.7, 20, and 30 au \citep[see, e.g.,][]{Pinilla2017}. 

We inject planetesimals in an annulus of width $\eta R_\text{ice}$ around the snow line following \cite{liu2019growth} and \cite{jang2022} with the planetesimals' mass function (PMF) drawn from streaming instability simulations \citep{Schafer2017}. This PMF features a characteristic planetesimal mass ($M_\text{char}$) and an exponential decay in the occurrence rate of large planetesimals. We note that $M_\text{char}$ varies with the local disk environment \citep[e.g.,][]{jang2022} and equals $1.6e^{-5}M_\oplus$ in our water-ice snow line models. We sampled the full PMF with Monte Carlo simulations constrained by the total planetesimal mass taken from \cite{liu2019growth}. However, limited by computational costs, we only included in the simulations planetesimals more massive than $3M_\text{char}$ (about 150 planetesimals) and discarded the numerous low-mass planetesimals \citep[see also][]{jang2022}. They make up half the mass of the whole planetesimal population \citep[see][Figure 5]{liu2019growth}. We shall see later that mutual collisions are limited for even these massive planetesimals, so most of them accrete independently. The ratio between the accreted pebble mass, $M_\text{peb}$, and the total planetesimal mass, $M_\text{plt,0}$, measures the pebble accretion efficiency for the whole population. 

The initial dust concentration is set as a Gaussian bump (centered around the snow line; see Figure \ref{fig:illustration}) imposed on a base value of $C_0=0.01$ whose peak and dispersion are $C_1=0.2$ and $\sigma=0.5$ au by default. Here $C_1$ reflects the amount of leftover solids not converted to planetesimals after the streaming instability \citep{Drkazkowska2017}. We vary the peak value in simulations in DustCon-1/2 of Table \ref{table:simulations} to study its influence. We also tried a more focused dust enhancement in the simulation FocusCon. The viscous alpha is constant, $\alpha=0.001$ or 0.0001 (Test1 model in Table \ref{table:simulations}), and the dust diffusivity is equal to the viscosity by default \citep{Zhou2022}. We investigate the effect of dust diffusion by turning off the dust diffusion or doubling the diffusivity in models DustDiff-1/2.

To save the computational cost, we truncate the disk at $R_\text{in}=0.1$ and $R_\text{out}$ = 40 or 80 au where absorbing boundary conditions are applied \citep{Zhou2022}. The initial value problem governed by the equations in Section \ref{sec:flux} is solved by the \emph{SciPy} package \citep{Scipy2020} with a fourth-order Runge-Kutta method on log uniform grids ranging from $R_\text{in}$ to $R_\text{out}$. We used 600 grids by default, and tests with 1200 grids show converged results. The surface density of gas and dust drops quickly near the inner edge ($<$ 2 $R_\text{in}$) due to the absorbing boundary condition. However, the boundary barely affects the mass growth of planetesimals near snow lines (see the Appendix \ref{sec:boundary}) except in simulation SnowLine-2, where we further set $R_\text{out}=80$ au.

We summarize the simulations above in Table \ref{table:simulations}. In addition, we carried out a simulation, Test2, similar to Fiducial, explicitly adding one massive planetesimal of mass 0.01 $M_\oplus$. We further investigated the effects of gravitational perturbations from Jupiter on the planetesimal accretion by running Test3. In this simple test, Jupiter's orbit is held fixed, and it does not affect the dusty disk evolution. 





\section{Results} \label{sec:results}

In this section, we overview the dust evolution and then the associated variation in pebble fluxes. The accretion of planetesimals is compared in different models. In all models, we found a decreasing planetesimal growth rate with time due to a decaying pebble supply. This is in contrast to previous studies with a constant pebble flux, where planetesimal growth accelerates.

\begin{figure*}[ht!]
\centering
\includegraphics[scale=0.6]{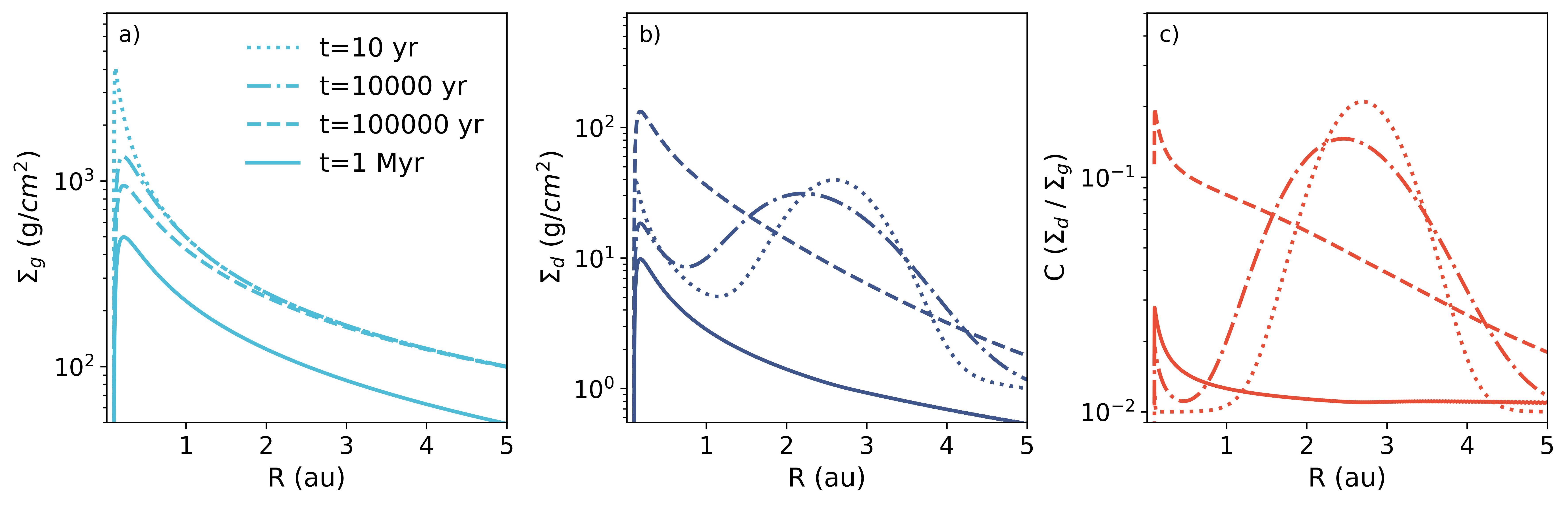}
\caption{Evolution of the gas surface density ($\Sigma_g$), dust surface density ($\Sigma_d$), and dust-to-gas ratio ($\Sigma_d$/$\Sigma_g$) in the early stage of the Fiducial simulation at different epochs.\label{fig:R-Sigma}}
\end{figure*}

\subsection{Dusty Disk Evolution}
In Figure \ref{fig:R-Sigma}, we show the evolution of the dusty disk in the Fiducial simulation, where the viscosity parameter $\alpha$ and the diffusivity coefficient $D_0$ are both 0.001. The initial dust concentrates around the snow line at 2.7 au, and the peak of dust concentration ($C$) is 0.2 plus the base dust fraction, 0.01. In the inner part of the disk, the gas keeps moving inward and being absorbed by the star. Dust is advected inward by the gas inflow while, in the meantime, the dust concentration profile broadens due to diffusion; moderate upstream dust diffusion is observed in the dust density \citep[see also][]{Zhou2022}.  Planetesimals placed around the snow line further accrete a small amount of dust, leaving no observable features in the dust profiles here. After around 0.1 Myr, the initial dust overabundance is erased, and a power law-like $\Sigma_d$ profile is established. The dust density then decreases with time throughout the disk. 

Variations in the initial dust concentration, as explored in the DustCon-1/2 simulations, lead to no qualitative changes. Doubling the dust diffusivity in DustDiff-2 also shows no significant influence except for stronger diffusion for the dust overabundance within 0.1 Myr.
However, when the dust diffusion is ignored in simulation DustDiff-1, the initial excess in dust concentration is advected inward, maintaining its shape, and still completely erased after 1 Myr. When the dust enhancement is placed further out (SnowLine-1/2) than the Fiducial run, a larger fraction of dust is transported outward due to gas advection (see Figure \ref{fig:pebbleflux}).


\subsection{Pebble Flux}

\begin{figure*}[ht!]
\centering
\includegraphics[scale=0.6]{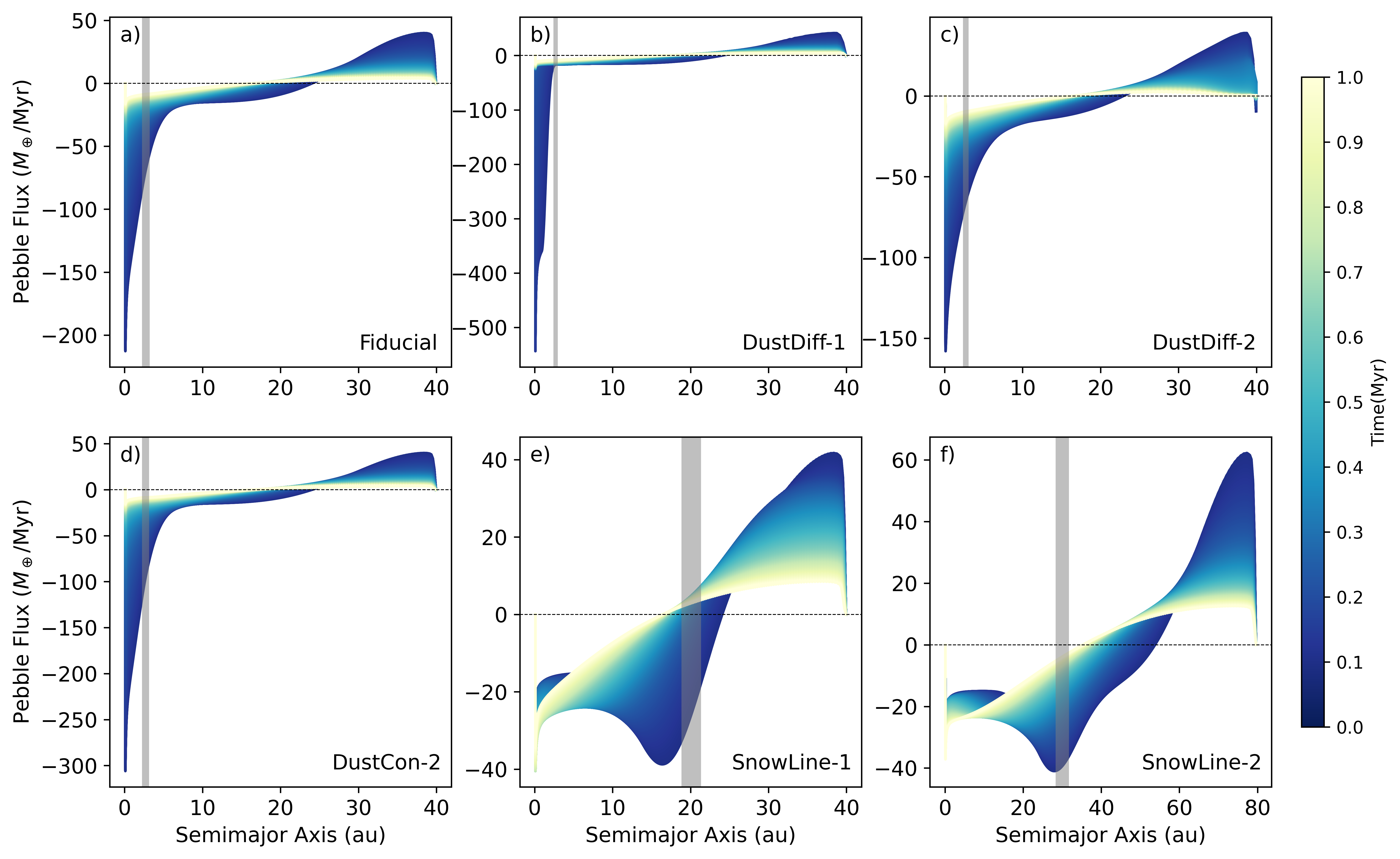}
\caption{Pebble flux given in Earth masses per Myr within the first Myr disk evolution (the time is color coded) in six models\label{fig:pebbleflux}. The grey shaded regions mark the planetesimals' semimajor-axis distribution by 1 Myr.}
\end{figure*}

The supply of pebbles is often characterized by the pebble flux, which measures the mass of pebbles across a certain circumference per unit time \citep[e.g.,][]{Morbidelli2015,liu2019growth,Ogihara2020}. Pebble flux is a critical parameter determining the final outcome of planetesimal pebble accretion. \cite{Lambrechts2019flux} showed that high pebble flux leads to the formation of close-in super-Earths, while low pebble flux leads to the formation of terrestrial planets.
For simplicity, a coherent pebble flux constant in space is regularly adopted in pebble accretion studies. However, it is well known that dust evolution can lead to variable pebble distribution and thus pebble supply \citep{Birnstiel2010,Lenz2019, Stammler2022, Zhou2022}. In Figure \ref{fig:pebbleflux}, we plot  the evolving pebble flux for a reference, although it is not used in our pebble accretion treatment (see Section \ref{sec:pltacr}). 

First, we note that the pebble flux can be positive  beyond about 25 au in our models. This is primarily due to the outward gas motion carrying the small pebbles \citep{Lynden1974,Liu2022}, which is often overlooked in pebble accretion studies. In the inner disk, the initial dust overabundance (a leftover of the streaming instability) is quickly lost to the star within 0.1 Myr, as shown in Figure \ref{fig:R-Sigma}. This leads to a significant inward pebble flux above 100 $M_\oplus$ Myr$^{-1}$, which is typically assumed in pebble accretion simulations. However, after the passage of the initial dust peak, the pebble flux drops to about 20 $M_\oplus$ Myr$^{-1}$ and keeps decaying, as in panel (a) of Figure \ref{fig:pebbleflux}. 

An increase or decrease of the dust enhancement leads to higher or lower pebble flux at comparable times as the Fiducial model. Adjusting the width of the dust excess in FocusCon leads to no qualitative changes in the pebble flux. When the dust diffusion is turned off, the advection of the dust peak is not delayed by any diffusion, so that a more abrupt pebble flux decay is observed in DustDiff-1, and DustDiff-2, with enhanced diffusion, is the other way around. 

The evolution of pebble flux is more complex when the initial dust enhancement is placed at 20 and 30 au. First, the pebble flux in these models is much lower than in the Fiducial model due to the lower dust surface density ($\Sigma_d$). As mentioned above, the dust enhancement is placed near the separatrix of the inward and outward dust motion so that we also observe significant outward transport of the dust overabundance in the SnowLine-1/SnowLine-2 model. We note that the outer boundary is absorbing (see the Appendix \ref{sec:boundary}), so as that dust reaches the outer boundary will never drift back. This is somewhat counterintuitive, but it reflects that beyond a certain radius, the gas's outward motion can prevent the inward drift of small pebbles. This outer boundary is far enough and has no effect on models except SnowLine-2 (see the Appendix \ref{sec:boundary}), where we move the outer boundary further out to 80 au. 

The pebble flux is dynamic and closely related to the dust evolution. In general, the pebble flux considered here is lower than that in previous studies with prescribed pebble fluxes.

\begin{figure}[ht!]
\centering
\plotone{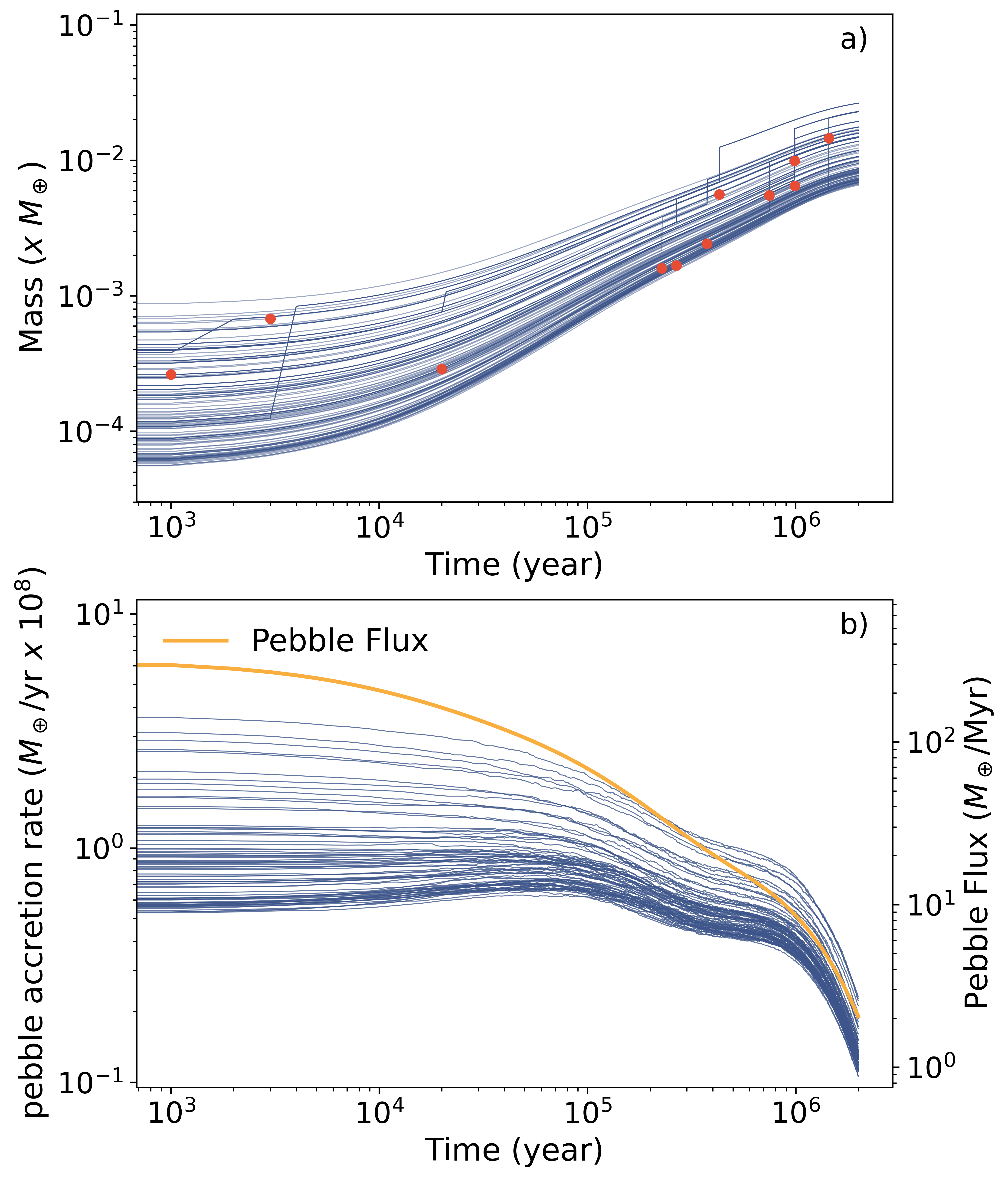}
\caption{Mass growth track (panel (a)) and pebble accretion rate (panel (b)) of Fiducial model. The red dots denote pairwise collisions between planetesimals that lead to the end of one mass growth track and a jump in mass for the other planetesimal. The yellow line denotes the absolute value of the pebble flux at 2.7 au.  \label{fig:t-Me}}
\end{figure}

\subsection{Mass Accretion }

Pebble accretion is believed to significantly boost the growth of planetesimals. \cite{liu2019growth} found that planetesimals can grow over an Earth mass within 1 Myr, assuming a constant pebble flux of 100 $M_\oplus/$ Myr$^{-1}$. However, the most massive object in our simulation is typically 2 orders of magnitude lighter than Earth, as summarized in Table \ref{table:simulations}. We note that \cite{liu2019growth} considered dust with $T_s=0.1$, which seems too large according to state-of-the-art coagulation models \citep[e.g., see][Figure 11]{Stammler2022}. Large dust particles are well settled in the midplane, leading to efficient  2D pebble accretion. On the other hand, they also drift easily and get lost to the star draining the pebble supply.  Therefore, it is not straightforward to assess the effect of dust size on pebble accretion. Nevertheless, recent studies found more efficient pebble accretion for smaller pebbles in the Bondi regime  \citep{Drakowska2021,Andama2022,lyra2023}. We will investigate this in depth in a following paper.

In Figure \ref{fig:t-Me}, we plot the mass evolution of planetesimals in the Fiducial simulation. First, collisions are few, so that the mass growth is mainly due to pebble accretion. The rarity of collisions is due to the small mutual Hill radius between planetesimals and also the effective damping of orbital excitation by the gas \citep[see also][]{liu2019growth}. The planetesimals capture pebbles of about 1 $M_\oplus$, 53 times their initial mass, after 2 Myr (see Table \ref{table:simulations}), signifying the importance of pebble accretion. However, our model is markedly different from previous studies in that the pebble accretion rate drops with time (Figure \ref{fig:t-Me}(b)). The pebble accretion is expected to become increasingly efficient as the planetesimals build up their mass, especially when the effective accretion radius is larger than the pebble scale height; i.e., 2D pebble accretion takes over 3D pebble accretion \citep{Lambrechts2014,liu2019growth,Ogihara2020}. 

\begin{figure}[ht!]
\plotone{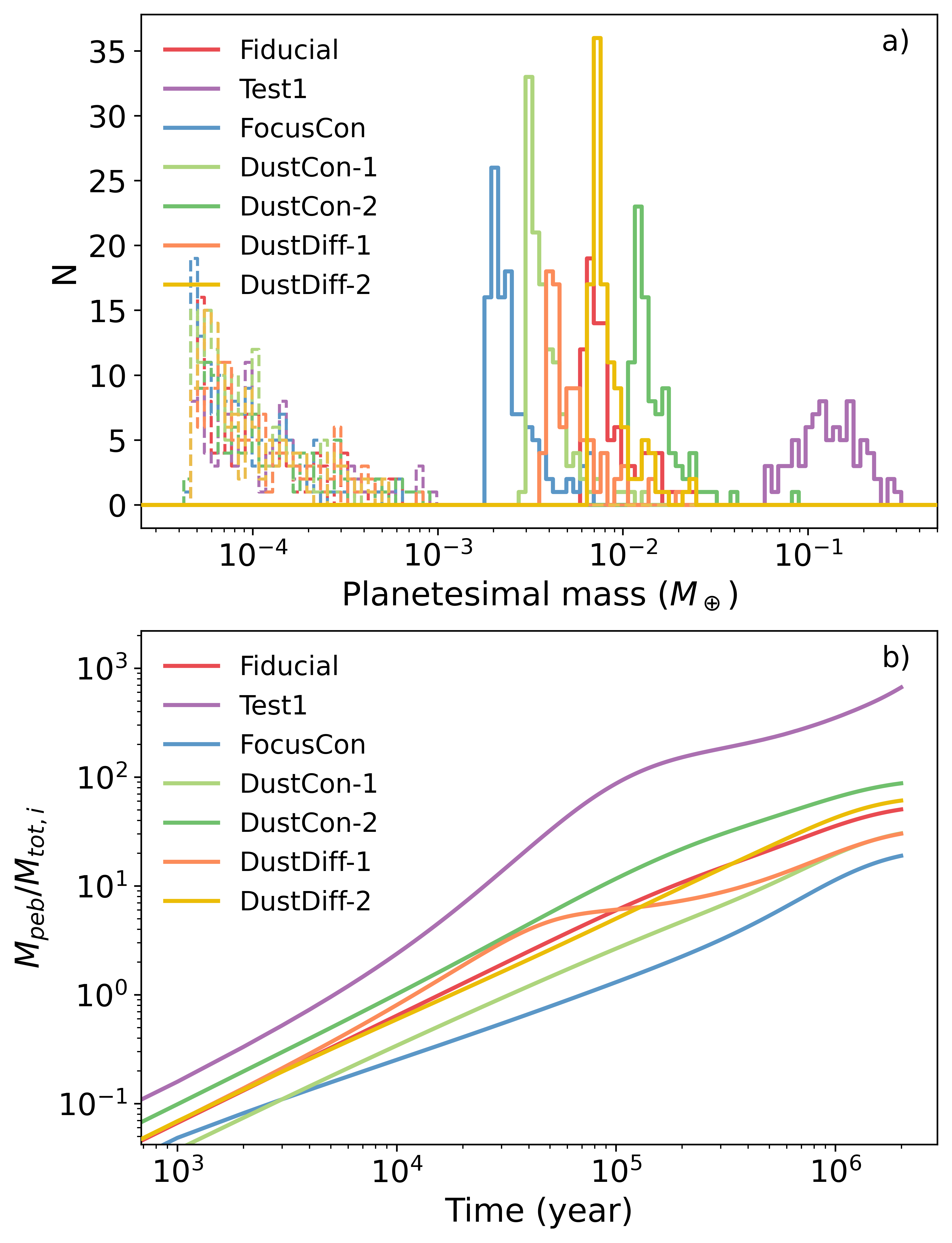}
\caption{Panel (a): the mass distribution of planetesimals at the beginning ($t_0$, the dashed lines) and in the end ($t_0$+2 Myr; the solid lines) in seven groups of models. Panel (b): the temporal evolution of the accreted pebble mass by all planetesimals \label{fig:mass_accreted}.}
\end{figure}

We attribute the decrease of pebble accretion efficiency in Figure \ref{fig:t-Me} to a decay in pebble supply, as they are correlated. As shown in Figure \ref{fig:R-Sigma}, the initial dust enhancement around the snow line is quickly lost to the star within 0.1 Myr, so that the pebble accretion rate starts to decrease. The pebble flux keeps decreasing with time, which limits pebble accretion after 1 Myr in Figure \ref{fig:t-Me}. The pebble accretion rate becomes so low, smaller than 0.1 $M_\oplus$ Myr$^{-1}$, that none of them can grow above a Mars mass within the lifetime of the gas disk based on simple timescale estimation. %

We plot the mass distribution of the planetesimals in different simulations in Figure \ref{fig:mass_accreted}(a). The population is dominated by planetesimals less massive than 0.01 $M_\oplus$ in the Fiducial simulation. Increasing the dust content immediately boosts the mass growth, with most embryos more massive than 0.01 $M_\oplus$, as observed in DustCon-2; decreasing the initial dust content naturally shifts the population toward lower masses as in DustCon-1 and FocusCon. The effect of dust diffusion is more subtle. As illustrated in Figure \ref{fig:pebbleflux}, high dust diffusion tends to delay the inward drift of dust and maintain a high pebble supply for a relatively long time so that the DustDiff-2 population shifts to the high-mass end. 

\begin{figure}[ht!]
\plotone{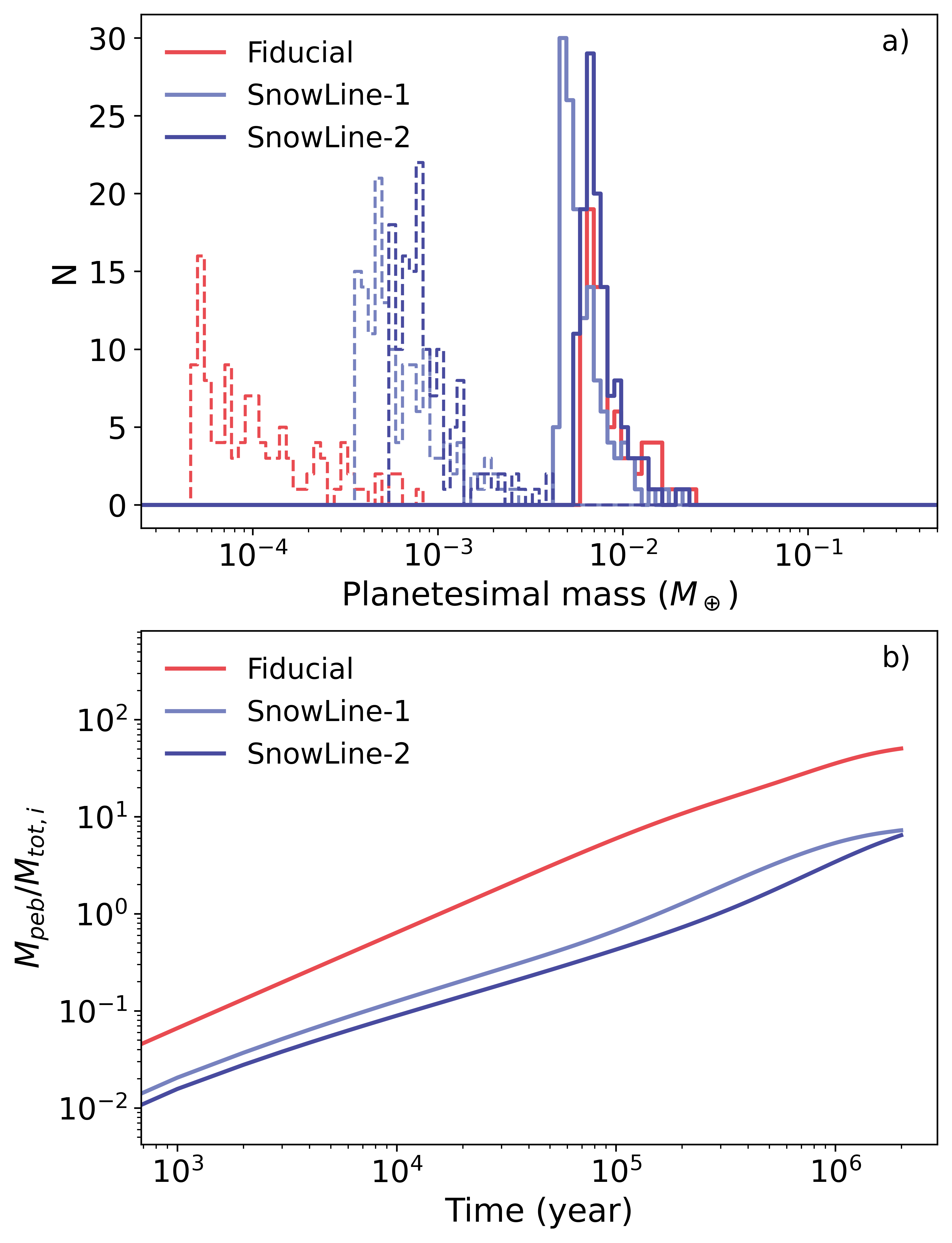}
\caption{Similar to Figure \ref{fig:mass_accreted} but for the SnowLine-1/2 models with larger initial planetesimals.\label{fig:SL12}}
\end{figure}

The Test1 model with low viscosity, $\alpha=0.0001$, stands out among the seven simulations, featuring more massive planet embryos and a flatter mass distribution. The reason is twofold. First, the gas accretion and the dust inward drift are slower in low-viscosity disks; second, the pebble scale height is smaller, and the efficient 2D pebble accretion starts early. Either way, the most massive protoplanets in Test1 are still less than 0.5 $M_\oplus$.

We find the ratio between the accreted pebble mass and the total planetesimal mass to be a good indicator of the pebble accretion capability, as plotted in Figure \ref{fig:mass_accreted}(b) (see also Table \ref{table:simulations}). By 2 Myr, it reaches about 50 in most simulations; i.e., 1 $M_\oplus$ pebbles are caught by the planetesimals of 0.018 $M_\oplus$. If more planetesimals are initially injected, the accreted pebble mass is expected to increase proportionally, since their growth tracks are almost independent of each other. The ring of planetesimals/protoplanets is expected to be excited by mutual gravity and collide to form planets after the gas disk is dispersed \citep[see, e.g.,][]{Hansen2009,Fang2020}, which is outside of the scope of the current paper on the pebble supply. 

In Figure \ref{fig:SL12}, we compare the SnowLine-1/2 models to Fiducial. At larger radii, the characteristic mass of the planetesimals is larger \citep[see, e.g.,][equation 6]{jang2022}, so the planetesimals are more massive than those in Fiducial. However, the pebble supply is also low at large radii, as is evident in Figure \ref{fig:pebbleflux}. Eventually, the growth of the planetesimals is limited by the pebble supply and reaches a final mass comparable to Fiducial at the end of the simulation. Their pebble accretion efficiency is even lower than Fiducial in Figure \ref{fig:SL12}(b).

\section{Discussion}
\label{sec:discussion}

The inefficient growth of planetesimals here is mainly due to a low pebble supply. Specifically, planetesimals formed by the streaming instability are light in mass, and they accrete slowly via 3D pebble accretion (Hill radius smaller than $H_\text{peb}$). When they become massive enough to accrete more efficiently via 2D pebble accretion, the supply of pebbles has decayed significantly. The mismatch of pebble supply and accretion efficiency suppresses the eventual pebble accretion and mass growth. 

\begin{figure}[ht!]
\plotone{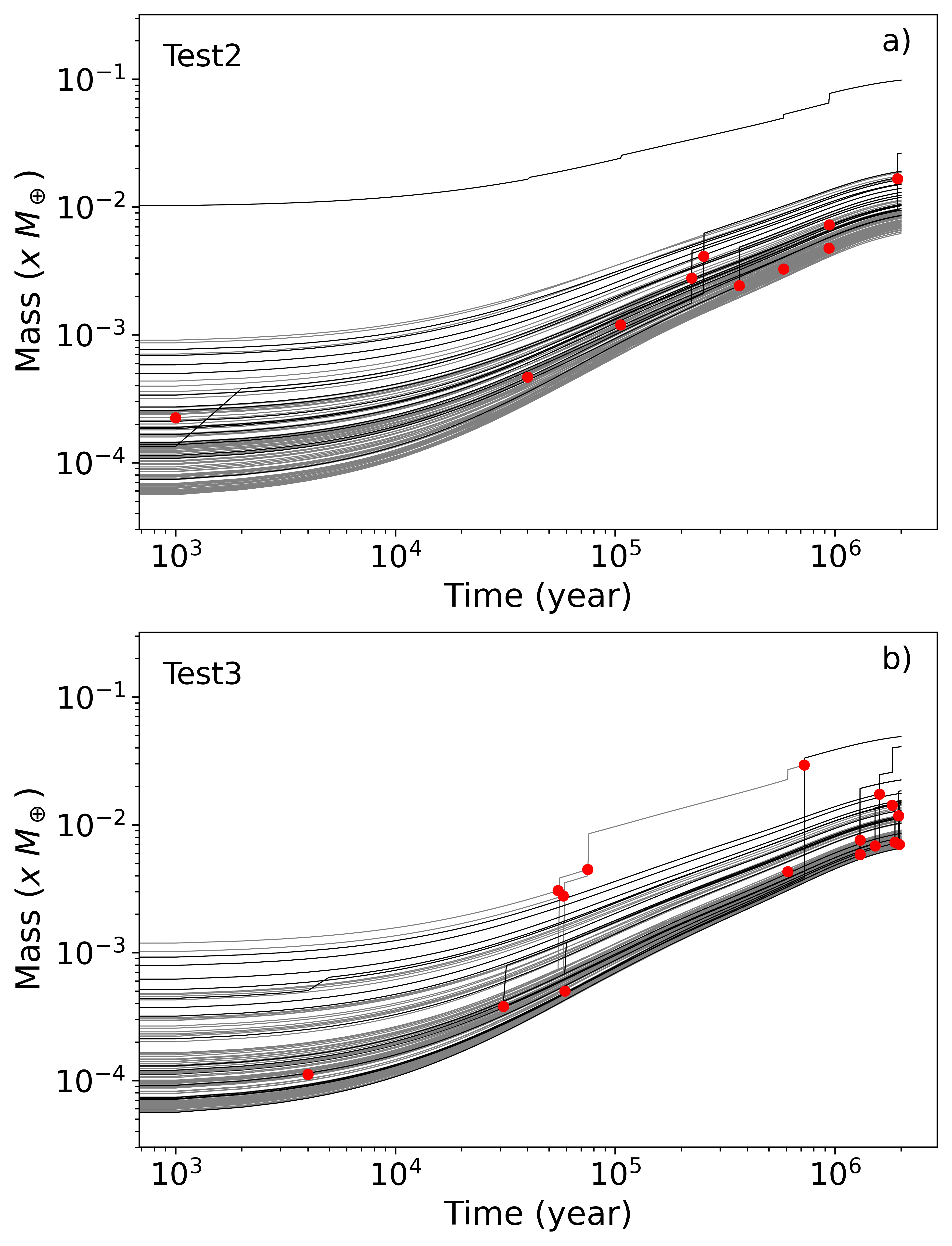}
\caption{Panel (a). the mass growth tracks of the Test2 model with an initial massive planetesimal. Panel (b). the mass growth tracks of the Test3 model, whose planetesimals are perturbed by the gravity of Jupiter. \label{fig:mass_Test2}}
\end{figure}

A natural way to circumvent the problem is to halt pebble drift and loss by disk substructures \citep{Lau2022}. However, it is still uncertain how the substructures form in the first place and whether they are produced by giant planets. According to our SnowLine-1/2 model, it is extremely challenging to grow giant planet cores at the large radii where substructures are observed. 
On the other hand, the early high pebble supply can be made full use of if massive planetesimals are present or produced. We thus carried out the Test2 simulation by including a massive planetesimal and the Test3 simulation by considering the gravitational perturbations of Jupiter. We plot their mass growth tracks in Figure \ref{fig:mass_Test2}.

More massive planetesimals may form via other disk instabilities than the streaming instability, e.g., the secular gravitational instability \citep{Pierens2021} and the collapse of pebble clouds in gravitationally unstable early disks \citep{Baehr2022}. We carried out a similar simulation to Fiducial and inserted a massive planetesimal of mass 0.01 $M_\oplus$ at the beginning. In Figure \ref{fig:mass_Test2}, the massive one does accrete more pebbles than its fellows. However, its final mass is still much lighter than previous models with a constant pebble flux \citep{liu2019growth}. 

Massive planetesimals can be produced via collisions and then speed up the pebble accretion. We thus place Jupiter planet in orbit to perturb and excite the orbits of the ring of planetesimals in the Test3 model. We note that Jupiter's effects on the dusty disk are ignored for simplicity. After 2 Myr, the maximum eccentricity of the planetesimals reaches 0.8 in the Test3 simulation, while in other simulations, the maximum eccentricity is less than 0.1. Indeed, collisions are more frequent, as shown in Figure \ref{fig:mass_Test2}, and the merged planetesimals lead the mass growth. 

Still, in these two tests, the growth of the massive bodies slows down due to the pebble supply decay. They do manage to reach a Mars mass after spending 2 Myr, but their pebble accretion rates are too low to allow them to independently grow above an Earth mass within the disk lifetime (see, e.g., Figure \ref{fig:t-Me}(b)). Accretion via collisions, albeit slow, may help them to grow further in mass. Many previous studies of pebble accretion start from planetary embryos of mass comparable to the most massive objects at the end of our simulations \citep[see, e.g.][]{Morbidelli2020,Ogihara2020,Jiang2022} and feed them with a pebble flux as large as 100 $M_\oplus$ Myr$^{-1}$. However, we note that by the formation of such embryos, the pebbles are significantly depleted if the disk bears no substructures.  Simulations of planetesimal growth \citep{liu2019growth,jang2022} all the way to the late giant impacts ($>10$ Myr) are desirable to assess planet formation with a dynamic pebble supply. 

Besides, we assume the initial disk mass of ${\approx}0.01 \ M_\odot$ in our fiducial model. This default setup aims at investigating the planetesimal growth and disk evolution in the class II disk phase. Nevertheless, the outcome can differ if the planetesimals form at a much earlier class 0/I phase \citep{Baehr2022}. A more massive reservoir of pebbles is present in such earlier disk phases \citep{jang2022}. As shown in section \ref{sec:flux},  $\Sigma_d$ increases proportionally to the initial gas surface density. If we consider a 0.1$M_\odot$ disk, the pebble accretion rate will increase by 10 times, and more massive planetesimals will form. More collisions are likely among more massive planetesimals, leading to a further boost in the accretion efficiency. However, such a disk already suffers gravitational instability and may experience very different dust evolution \citep{Rice2004,Zhou2022}.

\section{Summary}
\label{sec:conclusions}
We studied the planetesimal accretion in an evolving dusty protoplanetary disk by merging $N$-body simulations with the 1D semianalytical dust advection-diffusion model. We find a quick pebble loss to the star and variable pebble flux with space and time. Our pebble flux is on the lower end of the pebble flux assumed in previous studies and keeps decreasing with time.  Due to the low pebble supply, most planetesimals are stranded at a mass of less than a Mars mass after 2 Myr. Large planetesimals in low-viscosity disks are favorable for mass growth, but they still cannot grow above an Earth mass in our model. We find severe constraints on the pebble accretion efficiency due to a limited pebble supply. However, our models dealt with only one representative dust species. Therefore, studying polydisperse pebble accretion with a self-consistent pebble supply in various environments is highly desirable.

We thank Masahiro Ogihara for stimulating discussions. This work made use of the High Performance Computing Resource in the Core Facility for Advanced Research Computing at Shanghai Astronomical Observatory. H.Z. is supported by National Natural Science Foundation of China (NSFC; grant No. 12073010). B.L. is supported by National Natural Science Foundation of China (NSFC; grant No. 12222303, 12173035 and 12147103) and the Fundamental Research Funds for the Central Universities (2022-KYY-506107-0001,226-2022-00216). H.D. acknowledges support from the Chinese Academy of Science talent program and the Shanghai talent program. We thank the anonymous referee for comments that significantly improved the paper's clarity.
\software{Rebound \citep{Rein2012}, Reboundx \citep{Tamayo2020}, SciPy \citep{Scipy2020}}

\appendix
\section{The Effect of the Boundary Condition}
\label{sec:boundary}
\begin{figure}[ht!]
\plotone{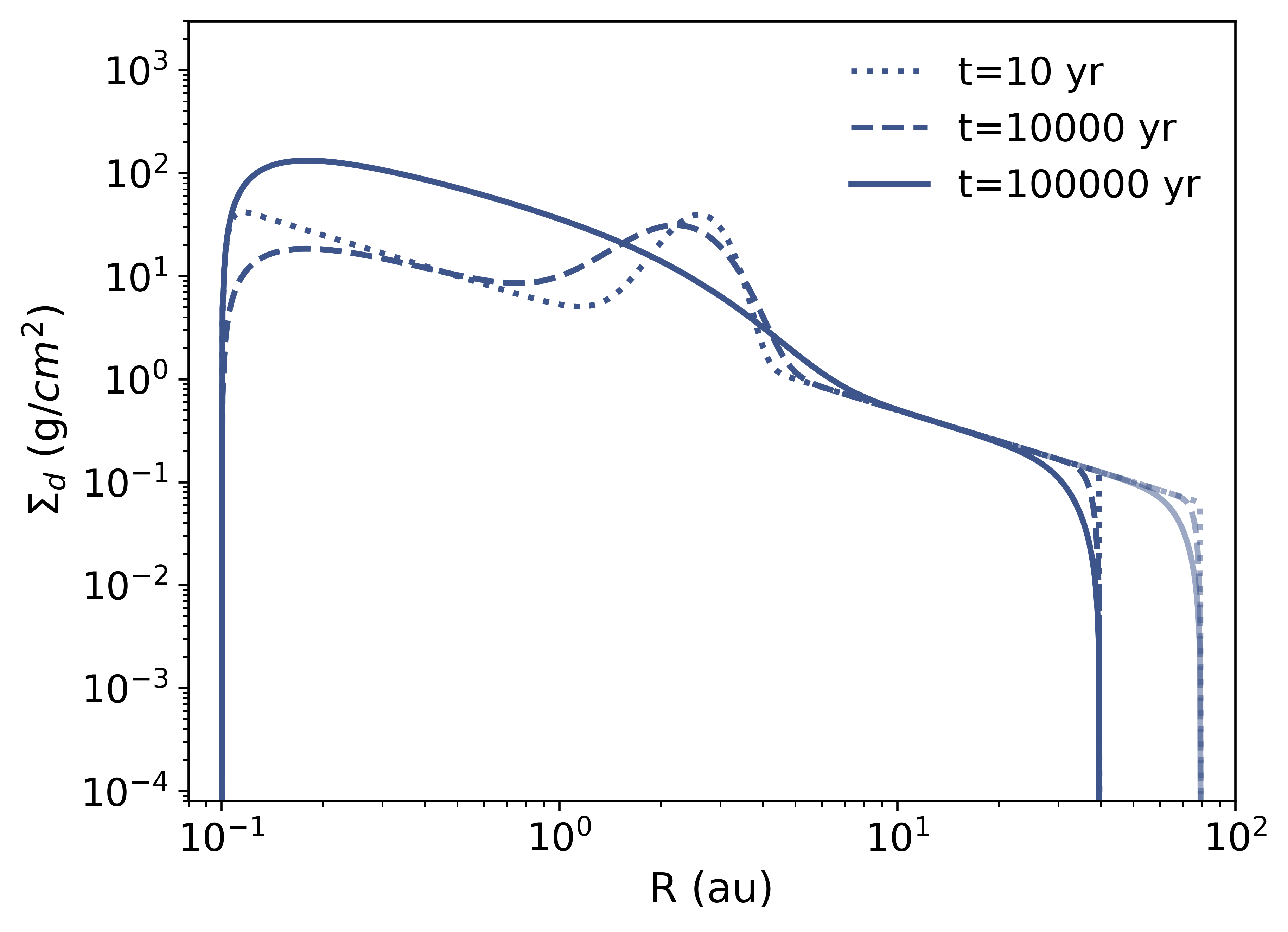}
\caption{Dust evolution in the Fiducial model with an outer edge at 40 and 80 au (lighter color lines).\label{fig:2Rout}}
\end{figure}

We adopted an absorbing boundary condition for the gas surface density by forcing its boundary value to be zero. The gas surface density evolves according to a diffusion equation so that the gas surface density smoothly transits to zero near the edge (within 2$R_\text{in}$). For the dust concentration, we extrapolate from the bulk of the computational domain to set its boundary values. To test the effect of the outer boundary, we rerun Fiducial by placing the outer boundary at 80 au. In Figure \ref{fig:2Rout}, the two models with different outer boundaries overlap in the region $R<20$ au. It is thus evident that our simulations of planetesimal growth near the snow line ($<20$ au) are not affected by the outer boundary, except in the SnowLine-2 model. We set $R_\text{out}=80$ au for the SnowLine-2 simulation to avoid boundary effects.


\bibliography{references}{}
\bibliographystyle{aasjournal}



\end{document}